# Meta-creatures: Developing an omnipotent hydrogel cell to construct bio-inspired systems

Hanqing Dai[1*†], Wenqing Dai[2†], Yuanyuan Chen[3], Wanlu Zhang[3], Yimeng Wang[5], Ruiqian Guo[1,3*], Guoqi Zhang[4*]

[1]*Academy for Engineering and Technology, Fudan University, Shanghai 200433, China.*
[2]*School of Materials Science and Engineering, Shanghai Jiao Tong University, Shanghai, 200240, China.*
[3]*Institute for Electric Light Sources, Fudan University, Shanghai 200433, China.*
[4]*Department of Microelectronics, Delft University of Technology, Delft 2628 CD, Netherlands.*
[5]*School of Science and Engineering, University of Dundee, DD1 4HN, Dundee, United Kingdom.*

[†]These two authors contributed equally to this work.
Corresponding author email: daihq@fudan.edu.cn; rqguo@fudan.edu.cn; G.Q.Zhang@tudelft.nl

**Due to current technological challenges, including the complexity of precise control, low long-term survival and success rates, difficulty in maintaining function over extended periods, and high energy consumption, the construction of life-like creatures with multilevel structures and varied physiological characteristics is a goal that has yet to be achieved[1-3]. Here, to create a parallel entity termed a meta-creature with similar functions and characteristics to those of natural organisms, we introduce a process for the transformation of an omnipotent hydrogel cell (OHC), which is inspired by totipotent stem cells and carries bio-inspired bioelectricity, into a meta-creature. The proposed OHCs demonstrate living-like behaviours such as proliferation, regeneration, and even micron-centimetre disintegration/reconfiguration; due to these behaviours, the OHC-based meta-creatures will be able to calmly handle numerous challenges affecting their survival and even revival from the dead state at a critical juncture. Moreover, we engineered bio-inspired bioelectricity-driven meta-nerve fibres, meta-skins, meta-mouths, and meta-cardiovascular systems. We captured the electrical signals transmitted between the meta-nerve fibres and rat sciatic nerves, the physicochemical signals perceived by the meta-mouth, and the information exchanged between the meta-skin and the external environment. Notably, the meta-cardiovascular system, which is capable of exchanging matter and energy with external environments, exhibited similar electrocardiogram signals during testing in rabbits; these results indicated feedback with a biological system and the potential for ex vivo bioelectric remodelling. Finally, a meta-creature was designed and exhibited bio-inspired bioelectricity signals during the simulated outdoor flight. These results indicated that the blood vessels, heart, and associated tissues within the body could be replaced or repaired when needed. This work reveals new possibilities for constructing bio-inspired systems, thereby improving our understanding of bioelectricity and biomimicry.**

# Introduction

The development of life-like bio-inspired artificial systems with multilevel structures and various physiological characteristics involves several advanced techniques, such as organoids and life-like systems[4]; these include pluripotent stem cell differentiation[5], the use of biomaterials through tissue engineering and 3D bioprinting[6], gene editing technologies[7], de novo synthetic biology[8], and bio-inspired electronic systems[9]. Notable advancements, such as the successful construction of various organoids (e.g., brain, gut, and liver)[10,11] and the creation of complex tissues (e.g., functional blood vessels, skin, and cartilage tissues)[12,13] through 3D bioprinting, highlight significant progress in disease modelling, drug testing, and regenerative medicine. In synthetic biology, Craig Venter's creation of a synthetic cell capable of self-replication has demonstrated the potential for designing genetic characteristics[14]. Additionally, bio-inspired electronic systems such as artificial skin[15,16] and eyes[17,18] enhance human–robot interactions and improve the quality of life of patients. These methods and advancements not only advance our understanding of the essence of life but also provide promising prospects for future biomedical applications.

Thus far, bio-inspired artificial systems are broadly classified into bio-inspired electronic systems and bio-inspired bio-hybrid systems. Bio-inspired electronic systems consist of non-living components such as sensors[19], biomaterials[20] and integrated circuits[21]; these mimic the biological functions and provide innovative solutions in fields such as soft robotics, tissue engineering, and medical diagnostics. These systems provide advantages in efficiency, biocompatibility, and sustainability. In contrast, bio-inspired bio-hybrid systems combine living and non-living components to perform functions such as gene expression[22], enzyme catalysis[23], heredity[24], metabolism[25], and biological regulation[26]. These systems are more compatible with living tissues, allowing for self-repair and more effective adaption to environmental stimuli than purely mechanical systems; these systems show significant promise in robotics, medical devices, and adaptive materials.

Despite these advancements, challenges remain in constructing life-like bio-inspired artificial systems with multilevel structures and physiological characteristics. For bio-inspired electronic systems, a large number of non-living components are needed to mimic functions such as molecular penetration and muscle contraction[27]. This typically requires highly efficient, low-power or even zero-power sensors to prevent overheating and ensure reliable operation. These sensors need to function harmoniously within biological environments, mimicking the energy efficiency of natural systems and minimizing their ecological impact. For bio-inspired bio-hybrid systems, the methodological challenges include achieving sufficient compositional diversity, chemical complementarity, systemic integration, and functional complexity[28]. Ensuring comparable functionality between the biological and nonbiological components is difficult due to the stringent requirements for safety, longevity, biocompatibility, and sensitivity. Additionally, the development of life-like bio-inspired artificial systems faces other overarching challenges, including suitable biomaterials, interdisciplinary collaboration, ethical considerations and scalability[29,30]. These challenges limit the functional diversity, structural reconfigurability, and information interaction capabilities of bio-inspired devices, hindering the creation of bionic devices based on biological growth and development processes. Consequently, constructing meta-creatures, which are life-like parallel entities with functions and characteristics similar to those of natural organisms, has yet to be achieved and is hindered by the intricate interplay of the biological and artificial components.

To address the aforementioned challenges and obtain a meta-creature, we utilized an omnipotent hydrogel cell (OHC) via free radical polymerization as an electrified meta-cell to develop meta-nerve fibres, a meta-skin, a meta-mouth, a meta-cardiovascular system, and a meta-creature. Moreover, animal experiments revealed the signal transmission ability of meta-nerve fibres and the ex vivo bioelectric remodelling ability of the meta-cardiovascular system. Furthermore, we thoroughly investigated the mechanisms of ion-selective diffusion, material transport and bio-inspired bioelectricity through molecular dynamics simulations and finite element simulations, respectively. Ultimately, our designed meta-creature with bio-inspired bioelectricity can facilitate new possibilities for constructing complete life-like parallel entities. These findings will provide new methods for implantable devices, prosthetics, wearable health monitors, soft robotics, brain–computer interfaces, and even resurrection and immortality.

## Results

### Design principles and structure of the OHC

Natural organisms not only lack the aforementioned challenges but also excel in energy conservation since most of their life activities (touch, taste, heartbeat, etc.) require extremely weak currents and voltages, typically microampere-level currents and voltages ranging from microvolts to several hundred millivolts[31]. If a natural organism can be considered a bioelectric machine, its smallest functional unit is a charged cell. These cells carry electrical potentials across their membranes and communicate, metabolize and move through electrical signals to maintain the basic functions of organisms. Moreover, the germ cells of multicellular organisms are initially transformed into totipotent stem cells with differentiation potential. These totipotent stem cells have the potential to develop into any type of cell. With the development of organisms, these totipotent stem cells gradually differentiate into different cell types and then form various tissues and organs. Through this process, organisms can build complex structural and functional systems to adapt to their living environment. Using these characteristics of organisms, we have some possibilities for constructing complete life-like parallel entities.

Since hydrogels have emerged as exemplary imitations of living tissues owing to their exceptional water content and retention, unparalleled flexibility, eminent biocompatibility, and microstructures resembling the extracellular matrix[32], our proposed strategy involves the design of an OHC with current and voltage output capabilities similar to bioelectricity and with functional characteristics similar to those of totipotent stem cells. Hence, this OHC could serve as the fundamental functional and characteristic electrified unit for generating bio-inspired bioelectricity and constructing meta-creatures and their associated meta-tissues and meta-organs, such as meta-nerve fibres, meta-skins, meta-mouths, and meta-cardiovascular systems. In addition, we can use current and voltage as important indicators of bio-inspired bioelectricity to describe its characteristics and behaviour.

The generation of bioelectricity in cells originates from ion exchange processes across the cell membrane; these processes create a concentration gradient of $Na^+$ and $K^+$ between the intracellular and extracellular environments, thereby establishing the membrane potential. Based on this, we designed the OHC using bioelectricity and ion-selective diffusion according to the concentration differences. To enable OHCs to mimic the electrical signal generation mechanism of nerve cells through ion-selective diffusion to generate bio-inspired bioelectricity (**Fig. 1a**), we designed the OHC structure shown in **Fig. 1b**.

Since hydrogels have excellent biocompatibility and NaCl is more biologically friendly, the OHC is composed of a reduced graphene oxide composite with cation-selective and anion-selective hydrogels and two ionic hydrogels containing high and low concentrations of NaCl; these are denoted as (rGO)C, (rGO)A, H and L, respectively (**Fig. 1b**). An intrinsic theoretical limitation is that irreversible ion diffusion along the concentration gradient inevitably leads to the fact that OHCs designed based on ion diffusion cannot achieve a stable output of current and voltage simultaneously; thus, to overcome this intrinsic theoretical limitation, we initially chose rGO to enhance and reconcile the selective function of the ion-selective groups to simultaneously achieve a stable output of current and voltage.

Among the OHC components, free ions ($H^+$, $Na^+$, $OH^-$ and $Cl^-$) function similarly to neurotransmitters by generating electrical signals. The L hydrogel serves as a receptor, whereas the (rGO)C and (rGO)A hydrogels act as voltage-gated ion channels. Notably, the (rGO)C hydrogel, which is rich in cation-selective groups ($-SO_3^-$), facilitates the easy passage of cations. Conversely, the (rGO)A hydrogel, which has abundant anion-selective groups ($-N^+(CH_3)_3$), allows easy passage of anions. The OHC could mimic the electrical signal generation mechanism of neural cells by producing current and voltage through ion-selective diffusion; thus, bio-inspired bioelectricity is generated that could be used for the conversion, conduction, transmission and encoding of information in the meta-creatures.

**Bio-inspired bioelectricity and electrochemical properties of OHCs**

To validate the above design, the bio-inspired bioelectricity of OHCs was characterized using a Keithley 2400 Source Meter at room temperature (**Fig. 1c**). The results clearly demonstrated that the 25 mm long OHCs with a cross-sectional area of 0.785 cm$^2$ had comparable currents (-31.11 ~ +15.37 μA) and voltages (-53.55 ~ +15.39 mV) with respect to those of humans and their organs or cells[33-36]; for example, the following data applies to humans: humans (42.6 ± 3.7 μA cm$^{-2}$, 10~100 mV), human corneas (0.07 μA cm$^{-2}$, 25~45 mV), human skin (1 μA mm$^{-1}$, 10~60 mV), skeletal muscle cells (-90 ~ +30 mV), smooth muscle cells (-60 ~ +10 mV), nerve cells (-70 ~ +40 mV) and cardiomyocytes (-90 ~ +20 mV) (**Supplementary Text 1**, **Table S1**). These results indicated that the OHC could serve as the minimal functional and characteristic unit for constructing meta-creatures and their associated meta-tissues and meta-organs. In addition, the electrophysiological properties of meta-creatures composed of OHCs may be comparable to those of human cells.

Moreover, the bio-inspired bioelectricity of the OHC reveals that the OHC can achieve a maximum current and voltage of 15.37 μA and 15.39 mV, respectively. A relatively stable voltage of 14.68 mV was maintained over a duration of 5.46 hours. Moreover, the output current of the OHC was consistent over time, and a relatively stable current of 13.58 μA was sustained for 2.51 hours (**Fig. 1c**). Notably, the overlapping current–voltage curves demonstrate that the OHC could simultaneously generate relatively stable outputs of 14.68 mV and 13.58 μA for 2.51 hours; these results indicated that the durable and stable bio-inspired bioelectricity of the OHC could ensure the stability required for the physiological characteristics and life activities of the meta-creature.

**Mechanism of bio-inspired bioelectricity for OHCs**

To elucidate the mechanism of OHC output stabilization, we employed molecular dynamics simulations to study the kinetic mechanism of ion diffusion for the OHC (**Supplementary Text 2**). The simulation results demonstrate that the $Na^+$ ions from the H right hydrogel migrate towards the (rGO)C hydrogel, whereas the $Cl^-$ ions from the H left hydrogel migrate towards the (rGO)A hydrogel (**Supplementary Fig. S1**). Hydrone diffusion plays a major role in the H right-(rGO)C structure. Then, the $OH^-$ and $H_3O^+$ ions emerge as the predominant diffusion ions in the L-(rGO)C structure. Moreover, the selective diffusion of cations and anions can be modulated by the (rGO)A and (rGO)C hydrogels.

Furthermore, the simulations reveal that rGO adsorbs cations since the reported results indicate strong π–π interactions[37-39], and water molecules gather around the hydrogel's functional groups (**Supplementary Figs. S2-5**). The incorporation of rGO into the A hydrogel weakens its selectivity for anions, resulting in initially low current and voltage of the OHC since the ions from ion diffusion have not yet reacted with the hydrogel (**Fig. 1c**). Due to the strong interaction of the π-π bonds in rGO, the (rGO)C hydrogel enhances the cation selectivity, and the (rGO)A hydrogel weakens the anion selectivity, leading to changes in the current and voltage directions of the OHC after some time.

Additionally, the highly active sites in rGO adsorb some $Na^+$ and $Cl^-$ ions from the H hydrogels in contact with the (rGO)C and (rGO)A hydrogels (**Supplementary Figs. S2-5**), hindering ion diffusion over time. This eventually results in stable potential on both sides where H hydrogels contact the (rGO)C and (rGO)A hydrogels. Consequently, the OHC reaches a stable voltage output (**Fig. 1c**). When the diffusion rates of the free ions ($H^+$, $Na^+$, $OH^-$, and $Cl^-$) in the H hydrogels stabilize, the current also reaches a stable output. The voltage stabilizes earlier and remains stable longer than the current due to the thickness of the H hydrogel (**Fig. 1c**). Moreover, when the ion diffusion is unrestricted, the $Na^+$ ions continuously migrate towards the (rGO)C hydrogel, and the $Cl^-$ ions migrate towards the (rGO)A hydrogel. This eventually results in a low potential on the L hydrogel side in direct contact with the (rGO)C hydrogel (**Supplementary Fig. S5**) and a high potential on the side in direct contact with the (rGO)A hydrogel (**Supplementary Fig. S4**). Thus, incorporating rGO into the cation-selective and anion-selective hydrogels is effective for obtaining relatively stable outputs of bio-inspired bioelectricity in OHCs.

## Metabolism and regeneration of OHCs

Nevertheless, the ability of OHCs to generate a stable bio-inspired bioelectricity output based on ion concentration gradient diffusion cannot be sustained permanently. This is because OHCs do not have the capability of cells to utilize ATP energy to maintain the ion concentration gradient between the inside and outside the cell. Since we are currently unable to provide OHCs with vitality, OHCs lack the autonomous capability to procure raw materials. Therefore, until OHCs possess these autonomous abilities, all metabolic and regenerative processes will require our assistance. In this study, the metabolic, proliferative and regenerative processes are identical to the free radical polymerization process used to prepare OHCs.

Considering that each component unit of OHC is produced via in situ free radical polymerization, we can artificially assist OHC in synthesizing new components via this method within seconds after the mixed raw materials are acquired from the environment. This process involves eliminating components that have lost their discharge capacity and replacing them, thereby achieving metabolism and regeneration similar to those of biological systems. Similar to biological systems, OHC can maintain the necessary and sufficient ion concentration gradient required for generating bio-inspired bioelectricity via metabolism and regeneration.

To elucidate the metabolic and regenerative capabilities of OHC, we utilized a 30-degree triangular cutter to excise one-twelfth of the H, L, (rGO)C and (rGO)A hydrogels of the OHC (25 mm long and 0.785 cm$^2$ cross-sectional area) to create a gap (**Supplementary Text 3**). The excised portion was then repositioned to its original gap to measure its current–time relationship during OHC metabolism and regeneration. Initially, the current of the OHC was recorded at 8.03 µA. At 15.5 s into the current–time relationship test, the excised H hydrogel was removed using insulated tweezers, resulting in a measured current of 5.31 µA (**Supplementary Fig. S6a**). The partial removal of the H hydrogel at the (rGO)C junction led to a decrease in the anion content, causing a lower potential at this point and an increased potential difference across the tested structure, thereby accelerating ion diffusion. The reduced contact area between the H and (rGO)C hydrogels increased the resistance of the structure. Consequently, the current of the tested structure slightly increased from 5.31 µA to 5.91 µA after 15.5 s and then gradually decreased with increasing ion diffusion time. At 52.3 s, a prepared solution for H hydrogel synthesis was added to the H hydrogel gap. Approximately 3 s later, the excised portion regrew, and the current–time relationship of the OHC returned to its original trend.

This occurred because the L hydrogel in the OHC structure served to gather cations and anions passing through the (rGO)C and (rGO)A hydrogels. Although the newly generated portion at the H hydrogel gap provided additional ions to the tested structure, the capacity of the L hydrogel to gather cations and anions was limited. The L hydrogel's gathering capacity had already decreased before the new portion formed at the H hydrogel gap, resulting in a continued decrease in the OHC's current even after the new portion formed. Moreover, the same procedure was applied to the (rGO)C and (rGO)A hydrogels (**Supplementary Fig. S6b, d**), and both revealed similar trends.

However, when the same procedure was applied to the L hydrogel, the results were significantly different (**Supplementary Fig. S6c**). The current–time relationship of the OHC indicated that the initial current was 11.04 µA. When the excised L hydrogel was removed with insulated tweezers, before the excised L hydrogel regenerated, the observed pattern remained consistent; after the removal of the excised (rGO)C and (rGO)A hydrogels, trends were observed, with the current initially increasing from 5.18 µA to 6.14 µA and then gradually decreasing to 4.66 µA as the ion diffusion time increased. However, when the prepared L hydrogel solution was added dropwise onto the gap of the L hydrogel, the current of the measured structure increased from 4.66 μA to 9.31 μA within 7.9 s as the excised part regrew at the gap.

The role of the L hydrogel is to gather the cations and anions passing through the (rGO)C and (rGO)A hydrogels in the OHC structure. When the excised L hydrogel was removed, the contact area between the (rGO)C and L hydrogels and between the (rGO)A and L hydrogels decreased; this led to an increase in the potential difference in the "H-(rGO)C-L" and "L-(rGO)A-H" structures and an increase in the resistance within the structure. Although the ion diffusion rate increased, the current only slightly increased. When the excised part regrew at the L hydrogel gap, the resistance within the measured structure decreased. Furthermore, the newly grown part had a lower ion content than the excised part; this resulted in an increased potential difference in the "H-(rGO)C-L" and "L-(rGO)A-H" structures. These factors collectively caused a sharp increase in the current of the measured structure to 9.31 μA within a short period after the regrowth of the new part at the L hydrogel gap. Based on these results, the strategy of OHC obtaining raw materials from the environment and synthesizing new parts through radical polymerization to replace the original defective parts could partially restore some of the original capabilities of OHC. This process is analogous to how cells sustain vital activities and adapt to their environment through metabolism, proliferation, and regeneration.

**Proliferation and regeneration of OHCs**

Metabolism provides the energy and material basis necessary for cellular functions, and proliferation and regeneration ensure cellular renewal and the continuous growth and repair of organisms. As the smallest functional and characteristic unit of meta-creatures, OHC enables each component (H, L, (rGO) C and (rGO) A hydrogels) to independently output bio-inspired bioelectricity and allows the proliferation and regeneration of the entire H, L, (rGO)C and (rGO)A hydrogels before any of these components lose their potential for generating bio-inspired bioelectricity; these aspects are essential for maintaining the life activities and environmental adaptation of meta-creatures. These characteristics can even enable the revival of meta-creatures in desperate situations.

To verify whether OHC really has the above super abilities, the bio-inspired bioelectricity relationships of the H, L, (rGO)C and (rGO)A hydrogels (each hydrogel: 5 mm in length and 0.785 cm$^2$ in cross-sectional area) were initially determined within 8000 s (**Supplementary Text 4**). The results indicate that the H, L, (rGO)C and (rGO)A hydrogels can output currents of 0.14~5.05, 0.31~6.61, 0.06~4.53 and 2.86~48.83 μA, respectively (**Supplementary Figs. S7 and S8**). In addition, their output voltages are -3.63~6.75, 15.47~35.85, -3.82~39.52 and 37.88~125.25 mV, respectively. These results indicate that each of the OHC components are independently capable of generating bio-inspired bioelectricity, and these characteristics indicate that meta-creatures composed of OHCs have the potential to sustain life-like activities such as metabolism, proliferation, regeneration, disintegration and reconfiguration.

Afterwards, we tested the ability of the OHCs to assimilate raw materials to support their proliferation and regeneration functions. Since the temperature at which these raw materials are polymerized into hydrogels by free radical polymerization exceeds 40 °C, we introduced water-soluble thermochromic materials into the raw materials to visualize the generation of the H, (rGO)C, L and (rGO)A hydrogels (*detailed in* **Methods**). This thermochromic material appears as peacock green below 45 °C and then transitions to colourless when the temperature exceeds 45 °C.

The results revealed that when the raw materials for synthesizing the H, (rGO)C, L, (rGO)A and H hydrogels were sequentially incorporated into the empty glass shell of the OHC (*detailed in* **Methods**), they subsequently regenerated into new units based on the observed rate of the visible colour change (**Fig. 1d**). Significantly, the results reveal that all the regenerated structures, including H, H-(rGO)C, H-(rGO)C-L and H-(rGO)C-L-(rGO)A, exhibit operational capacities, and they can output initial currents of -2.64, 5.68, 6.99 and 1.53 μA, respectively. Thus, OHC still has the bio-inspired bioelectricity needed to maintain the life activities of the meta-creature during the regeneration of these hydrogel modules. Furthermore, our results reveal that the fully regenerated OHC successfully recovers its operational capacity with an output current of 5.63 µA (**Fig. 1d**), showing the robustness of the proliferation and regeneration process. When the newly generated OHC temperature returns to room temperature, its output current can reach 5.74 µA. This remarkable proliferative and regenerative ability could help support the development of meta-creatures with multilevel structures and even resuscitate meta-creatures on the verge of extinction, enabling them to address challenges that may impact their survival in natural environments.

## Macroscale disintegration and reconfiguration of OHCs

Furthermore, we noted that all the aforementioned H, (rGO)C, L, (rGO)A, H-(rGO)C, H-(rGO)C-L and H-(rGO)C-L-(rGO)A can exhibit good operational capacities; thus, they can be regarded as generalized OHCs, which are hydrogel cells capable of generating bio-inspired bioelectricity and possessing similar abilities for metabolism, proliferation, and regeneration as living creatures. Hence, the (rGO)C-L-(rGO)A, L-(rGO)A, H-(rGO)A-L, L-(rGO)C, H-(rGO)A, H-(rGO)C and H-(rGO)C-L structures can also be viewed as generalized OHCs, and their bio-inspired bioelectricity relationships indicate that their initial currents are 2.66, -3.03, 7.16, 8.98, 12.19, 7.39 and 8.09 µA, respectively (**Supplementary Text 5, Supplementary Figs. S9-15**). Furthermore, we speculate that even if OHCs disintegrate into segments, reconfiguring and reattaching those segments might enable them to regain their original functionality.

To test this hypothesis, we initially investigated the resilience of OHCs to macroscale disintegration and reconfiguration at room temperature (**Supplementary Movie S1**). Here, H, (rGO)C, L, and (rGO)A are all 10 mm in diameter and 5 mm in thickness. First, we started testing the bio-inspired bioelectricity of a complete OHC, and its initial current was 4.97 µA (**Fig. 2a**). When the current decreased to a low value (0.81 µA), we quickly removed the OHC from the testing box and disassembled it into individual H, (rGO)C, L, (rGO)A hydrogels. Immediately afterwards, each hydrogel was uniformly divided into four centimetre-size segments, and these segments were then randomly reassembled to form a new OHC as quickly as possible because the disintegration and reconfiguration of the OHC needed to be completed before any H, (rGO)C, L, or (rGO)A hydrogels lost their bio-inspired bioelectrical output. Here, only the operating subject had this disintegration and reconfiguration ability with the opportunity to use its remaining electrical power to reconfigure the disintegrated segments together in practical applications.

The results indicate that the operational capacity of the reconfigured OHC is successfully restored, and its initial output current is 7.77 µA (**Fig. 2a**); this value is more than 9 times greater than that of the previous 0.81 µA. Finally, the current of the reconfigured OHC plateaued at 1.75 µA; this value was more than double the previous 0.81 µA. These results not only confirm our hypothesis but also suggest that the macroscale disintegration and reconfiguration capability of OHC is easy to attain and apply and could be suitable for large structures, especially for developing meta-creatures or robots that can self-disassemble and reassemble, improving their flexibility and adaptability.

## Disruption and reconfiguration activities of the two OHCs

Here, we designed two nearly identical sets of disintegrated OHCs to explore the process by which they reconfigure into OHCs and release bio-inspired bioelectricity (**Supplementary Movie S2**). Each hydrogel sphere consisted of a 20 mm diameter magnetic sphere covered with a 3 mm thick layer of hydrogel. The segment of these hydrogel spheres connected to the black-striped wire was considered the tail, whereas the other end was considered the head. Under magnetic control, they sequentially self-reconstructed into two OHCs.

Furthermore, the results showed that the reconfigured OHC (OHC$_1$) that completed reconstruction at 108.37 s presented an initial current of 71.45 µA (**Fig. 2b**), whereas the OHC (OHC$_2$) that completed reconstruction at 193.31 s presented an initial current of 88.45 µA. The two reconfigured OHCs subsequently began to approach each other. They first connected head-to-tail at approximately 278 s, displaying bio-inspired bioelectricity currents of 37.03 μA. At 433 s, they separated head-to-tail, with $OHC_1$ exhibiting a current of 42.93 μA and $OHC_2$ showing a current of 31.49 μA. They then connected head-to-head at 496 seconds and tail-to-tail at 952 seconds. These findings indicated that a meta-creature or robot composed of OHCs could be applied in manufacturing to improve the flexibility and adaptability of the production lines.

**Microscale disintegration and reconfiguration of OHCs**

However, this macroscale disintegration and reconfiguration capability of OHCs may be difficult to apply to microstructures and microdevices for producing highly precise assembly and reconstruction. Fortunately, the significant increase in the output current of the reconfigured OHC provided the direction to solve this problem. If OHC can be reconfigured into micro-sized segments and the output current of the reconfigured OHC increases, microdevices, micro meta-creatures, microrobots and microbiological systems constructed from OHCs could have sufficient energy and be widely applied in fields such as microelectronics, microelectromechanical systems, biomedical engineering and life sciences.

We subsequently investigated the microscale disintegration and reconfiguration of the OHC at room temperature (**Supplementary Movie S3**). Unlike the previous options, we selected H, (rGO)C, L, (rGO)A hydrogels with 28 mm diameters and 16 mm thicknesses to easily obtain more OHC fragments for producing a more random reconstruction. Here, the diameter and length of the testing cylinder device are 28 and 85 mm, respectively. First, we started testing the bio-inspired bioelectricity of a complete OHC, and its initial current was 0.15 μA (**Fig. 2c**). When the current increased to 0.48 μA, we fragmented the OHC into micron-scale particles (**Supplementary Text 6, Supplementary Figs. S16-19**) with an electric crusher. These micron-scale particles were subsequently reconstituted in a random fashion to reconfigure a new OHC.

The results demonstrate that the operational capacity of the reconfigured OHC is restored, and its initial output current is 34.35 μA; this value is more than 71 times greater than that of the previous 0.48 µA (**Fig. 2c**). Herein, the 25 mm long OHC with a cross-sectional area of 0.785 cm$^2$ displayed a larger initial current than the OHC with a 28 mm diameter and 16 mm thick units, and their initial currents showed the opposite trend. The fundamental concept of electric current is defined as the quantity of electric charge that flows through a conductor's cross-sectional area within a given time interval. These results demonstrate the correctness of our problem-solving direction, indicating that OHC may also have great potential in the fields of precision engineering and micro devices.

**Series-parallel mechanism of disintegration and reconfiguration of OHCs**

To explain the above results, we proposed a series-parallel mechanism of miniature self-powered hydrogel cells (**Supplementary Text 7**). Specifically, due to the inherent independent power capabilities of the H, L, (rGO)C and (rGO)A hydrogels (**Supplementary Figs. S7** and **S8**), we hypothesize that after disintegration, each segment of these hydrogels can be regarded as miniature self-powered hydrogel cells with positive and negative electrodes. These miniature self-powered hydrogel cells can be freely combined into the aforementioned generalized OHCs with different structures, such as (rGO)C-L-(rGO)A, L-(rGO)A, H-(rGO)A-L, L-(rGO)C, H-(rGO)A, H-(rGO)C and H-(rGO)C-L structures (**Supplementary Figs. S9-15**). The reconfigured OHC can be viewed as series and parallel generalized OHCs composed of these microscale self-powered hydrogel cells (**Supplementary Fig. S20**).

Furthermore, the macroscale disintegration-reconfiguration results indicate that the initial current value (7.77 µA) of the reconfigured OHC is approximately 1.5 times greater than that of the original OHC (4.97 µA) and approximately 9 times greater than that of the pre-disassembled OHC (0.81 µA). Since these unbroken H, L, (rGO)C and (rGO)A hydrogels can output initial currents of 5.05, 6.61, 4.53 and 48.83 µA, respectively, these segments contained in the reconfigured OHC need to have a series-parallel result; otherwise, it violates this law stating that the current of the series circuit is determined by the largest output current component, and the current of the parallel circuit is the sum of the output current of all the parallel components.

Similarly, the microscale disintegrated and reconfigured OHC is expected to contain numerous series-parallel structures, causing the initial current value of the reconfigured OHC to be much greater than that of the OHC before reconfiguration. Clearly, the results show that the initial current value (34.35 μA) of the reconfigured OHC increases by 228 times compared with that before (0.15 μA), indicating that the reconfigured OHC contains many series-parallel structures. These results show that a greater number of monomers after disassembly correlates to a greater initial discharge intensity of the reconfigured unit. Thus, we may be able to temporarily improve the power output and hence the meta-creatures' attack and defence capabilities in a short period by disintegrating the OHC into many smaller pieces and reassembling these pieces into a new meta-creature, which may have considerable application potential.

## Ordered series and parallel connections for OHCs

Since the aforementioned disordered series–parallel connections of the disintegrated OHC can instantly increase its output capacity, the influence of the ordered series–parallel connections on the output capacity was investigated. This relationship is crucial for accurately and controllably maintaining the bio-inspired bioelectricity of bionic devices, human interface devices and meta-creatures composed of OHCs at a level equivalent to the bioelectricity of natural organisms. Moreover, this relationship can ensure the effectiveness of bionic organisms in terms of functional compatibility, biocompatibility, signal transmission, energy efficiency, and applications in research and therapy. A bioelectric level consistent with that of natural organisms promotes the normal operation of essential functions such as cellular activity and neural conduction, prevents immune responses and other rejection reactions, ensures proper electrical signal transmission, and enhances energy utilization efficiency. Additionally, this bioelectric matching is crucial for the reliability and effectiveness of medical research and therapeutic applications, such as in neural prostheses and cardiac pacemakers.

Initially, we connected OHCs in series in a cross-sectional manner, with the number of connections ranging from 2 to 9. The results illustrate that the initial voltages gradually increased, with values of 33.81, 50.85, 74.06, 61.37, 66.99, 79.50, 82.98 and 188.06 mV, respectively (**Fig. 3a**, **Supplementary Text 8**, **Supplementary Fig. S21**). Notably, their initial currents did not remain constant but increased almost linearly, with values of 9.80, 28.12, 35.57, 54.54, 72.50, 98.00, 119.71, 132.29 and 158.39 μA, respectively (**Supplementary Fig. S22**). Then, we connected 3, 5, 7, and 9 OHCs in parallel. The initial current gradually increased, with values of 14.4, 44.08, 58.99, and 68.95 μA, respectively (**Fig. 3b**, **Supplementary Fig. S23**). The initial voltages also gradually increased, with values of 24.64, 46.51, 55.05, and 74.81 mV, respectively (**Supplementary Fig. S24**).

After analysis, the answer to the anomalous phenomena observed in the series connection is hidden within the -(rGO)C-H-(rGO)A- structure of the series interface. According to the principle of diffusion driven by concentration gradients, the cations and anions in the H hydrogel diffuse into both the (rGO)C and (rGO)A hydrogels. This is distinct from the single OHC structure, where the H hydrogel diffuses only one type of ion into either the (rGO)C or (rGO)A hydrogel. Additionally, two hydrogels that are exactly identical are impossible to fabricate; thus, it equally impossible to create identical OHCs. Therefore, the initial output voltages of the OHCs in parallel connections are not uniform. This discrepancy leads to the observed phenomenon where the voltage increases with increasing number of OHCs connected in parallel.

Since OHCs rely on concentration gradient-driven ion diffusion to generate current and voltage, the parallel connection of these OHCs with different initial voltages does not result in circuit failure. Moreover, OHCs in parallel connections that exhibit capacitor-like characteristics can be present[40-42], causing an increase in the impedance within the circuit and a decrease in current flow. This capacitor-like characteristic causes difficult for OHCs to increase current in parallel. For example, we employed a configuration in which 18 OHCs were connected in an orderly manner in parallel, mimicking the structure of a bicycle chain (**Fig. 3c**, **Supplementary Movie**

S4). The results illustrate that with this ordered parallel connection, the 18 OHCs sustain a relatively stable current of approximately 14.50 μA over one hour (**Fig. 3d**), and the current does not increase. Moreover, this finding indicates that the stability of this ordered parallel connection is better than that of the aforementioned disordered series–parallel connections.

Afterwards, we connected 180 OHCs in series, resembling a bent bicycle chain (**Fig. 3e**, **Supplementary Movie S5**), forming a long snake-like structure on a tabletop. The advantage of this ordered series connection is the flexible rotation, ensuring that the overall system is both flexible and deformable. Despite the minimal contact area (0.1 cm$^2$) between the hydrogels and the limited free ion diffusion resulting from the series connections, the voltage output of approximately 5.11 V was consistently maintained for approximately 300 s via this ordered series connection comprising 180 OHCs (**Fig. 3f**). Once again, the results illustrate that the stability of this ordered series connection is better than that of the aforementioned disordered series–parallel connections.

Based on all of these results, while a large number of series connections can achieve very high voltages, they do not increase the output current and come at the cost of many OHCs. Compared with the series connection method, the disintegrated and reconfigured method can serve as a rapid and efficient alternative for enhancing output by using a relatively smaller amount of OHCs. Moreover, these results demonstrate that OHCs connected through ordered parallel and series can obtain currents and voltages comparable to those of natural creatures[36], organs[43-46], and implantable devices[47-51] (**Supplementary Text 9**, **Supplementary Table S2**). Moreover, these results indicated that OHCs may also have the ability to output bio-inspired bioelectricity when they develop into meta-tissues, meta-organs and meta-systems.

## OHC-based meta-tissues

To obtain a comprehensive validation of the ability of OHC-based meta-tissues to generate bio-inspired bioelectricity, we utilized a hosepipe as a mould for batch preparation of a 0.5-meter

**19**

long fibrous OHC as a bio-inspired nerve fibre meta-tissue (**Fig. 4a**, **b**) to investigate the electromechanical responses of this bio-inspired nerve fibre under various environmental stimuli, including twisting, dangling, stretching, and imitating rope skipping. The results reveal significant variations in current and voltage, and the untwisted bio-inspired nerve fibre achieves maximum values of approximately 0.58 μA and 0.46 V (**Fig. 4c**, **d**). These values follow a distinct curve, indicating a continuous variation in response to mechanical manipulation. In contrast, the twisted bio-inspired nerve fibre displays a fluctuating current and voltage with lower maxima of approximately 0.42 μA and 0.16 V (**Fig. 4c**, **d**), highlighting a unique electromechanical behaviour under torsion. The consistent fluctuation of current from an initial value of 0.43 μA when the twisted bio-inspired nerve fibre is suspended, coupled with its noticeable response to stretching and imitating rope skipping (**Fig. 4e**, **Supplementary Movie S6**), highlights the high sensitivity and energy conversion (transformation of mechanical energy into electricity) of the OHC-based bio-inspired nerve fibre to mechanical deformation. These findings not only validate the efficacy of OHC-based meta-tissues in generating bio-inspired bioelectricity but also show their potential for applications in bio-inspired bioelectricity harvesting and bioelectric sensing.

Since the mechanical responsiveness and adaptability to environmental changes make these OHC-based bio-inspired nerve fibre highly suitable for advanced bioelectronic systems, we explored their potential as nerve fibres to generate electrical signals in living animals. In these experiments, we initially used a voltage source to determine that the sciatic nerve of a rat can respond to stimulation at 99.4 mV (**Fig. 4f**) and then used a 0.5-meter bio-inspired nerve fibre to generate 123 mV bio-inspired bioelectricity to stimulate the sciatic nerve of a live rat to successfully induce leg movements (**Fig. 4g**, **Supplementary Movie S7**). This significant result indicates the feasibility of using bio-inspired nerve fibres as neuro-machine interfaces; thus, these could form the basis of the nervous system in meta-creatures. Such meta-creatures would be capable of responding to their environment through bioelectricity, facilitating exciting possibilities for future developments in bio-inspired robotics and synthetic biological systems.

Furthermore, to ensure that the length of the bio-inspired nerve fibre can span from the top to the bottom of a meta-creature, the OHC-based meta-tissue needs to be considered for scaling purposes. Since Gheorghe Muresan is the tallest basketball player in NBA history, stands at approximately 2.31 metres and does not exceed 2.4 metres, we again utilized a hosepipe as a mould for batch preparation of a 2.40-metre length bio-inspired nerve fibre (**Fig. 4h**); this length meets the needs of the exploration of large meta-organisms. The 2.40-m long bio-inspired nerve fibre can reach a voltage of 0.34 V (**Fig. 4i**, **Supplementary Movie S8**), and the voltage exhibits a characteristic pattern of initial increase followed by a subsequent decrease with increasing ion diffusion time. These results confirm that large-scale bio-inspired nerve fibres have the potential to serve as meta-nerve fibres that constitute the neural system for macroscale meta-creatures. These results also indicate that we can produce bio-inspired nerve fibres of substantial length and maintain functional voltage characteristics, showing the scalability of the technology; thus, these fibres are viable for creating complex and extensive neural networks in meta-creatures and facilitating the development of advanced bio-inspired robotic systems and synthetic biological entities.

## OHC-based meta-organs

To further demonstrate the versatility of OHCs in generating bio-inspired bioelectricity, we initially explored their application in sensory organs by manufacturing an OHC-based meta-mouth to distinguish among sour, sweet, bitter, peppery, and salty solutions. First, we utilized a mould to fabricate an OHC-based meta-mouth resembling the shape of a goldfish's mouth (**Supplementary Text 10**, **Supplementary Fig. S25**) and then connected it to a circuit to measure its bio-inspired bioelectricity signals. At approximately the 20-second mark, a sour solution with a concentration of 0.01 g mL$^{-1}$ was introduced to the OHC-based meta-mouth, the solution was held in its mouth for approximately 60 seconds and then the solution was expelled at the 80-second mark. Repeating these steps, we tested the bio-inspired bioelectric response signals of the OHC-based meta-mouth to sour solutions with concentrations of 0.02 and 0.03 g mL$^{-1}$. Additionally, we studied the bio-inspired bioelectric response signals of the OHC-based meta-mouth to sweet, bitter, pepper, and salty solutions with concentrations of 0.01, 0.02 and 0.03 g mL$^{-1}$, respectively.

Furthermore, the results indicate that the OHC-based meta-mouth exhibited an insensitive response to the 0.01 g mL$^{-1}$ sour solution (**Supplementary Fig. S26**). However, when sour solutions with concentrations of 0.02 and 0.03 g mL$^{-1}$ are ingested, significant and remarkably similar voltage gradient bio-inspired bioelectric response signals are observed. Moreover, the half-peak width of the voltage gradient bio-inspired bioelectric response signal for the 0.03 g mL$^{-1}$ sample was notably greater than that for the 0.02 g mL$^{-1}$ sample. Furthermore, we observed this phenomenon when the OHC-based meta-mouth was exposed to sweet, bitter, peppery, and salty solutions (**Supplementary Figs. S27-30**). The voltage gradient bio-inspired bioelectric response signals exhibited by the OHC-based meta-mouth show specificity when tasting these different flavours. These findings indicate that the OHC-based meta-mouth can not only detect the presence of sour, sweet, bitter, peppery, and salty stimuli but also differentiate these stimuli due to their varying intensities. This ability is indicative of a sophisticated level of sensory processing, similar to that of biological taste receptors.

Additionally, while the OHC-based meta-mouth was filled with 0.02 and 0.03 g mL$^{-1}$ solutions, the voltage gradient bio-inspired bioelectric response signals remained stable and consistent (**Supplementary Figs. S26-30**). In contrast, when these solutions were held at 0.01 g mL$^{-1}$, the OHC-based meta-mouth generated voltage gradient bio-inspired bioelectric response signals almost exclusively in the range of 60–65 s. After all solutions were expelled at the 80-second mark, only the 0.01 g mL$^{-1}$ sour solution at 112 s and the 0.02 g mL$^{-1}$ peppery solution at 118 s produced noticeable voltage gradient bio-inspired bioelectric response signals. Finally, to explore the practical application value of the OHC-based meta-mouth, we studied its ability to taste Nongfu Spring, Milk Deluxe, soya milk and chili oil, respectively (**Supplementary Figs. S31-34**). The results revealed that it showed differentiated bio-inspired bioelectric responses to them at each taste stage, such as intaking, tasting, suctioning and aftertaste states. These results indicate that meta-mouth responses closely mimic the neuroelectric signals associated with human taste perception. This finding is particularly exciting since it will facilitate possibilities for creating sensory organs in meta-creatures to experience and respond to environmental stimuli in a manner similar to that of humans.

To extend the applicability of the OHC, we designed an OHC-based meta-skin to illustrate its role in the movement and sensory perception of meta-creatures (**Supplementary Text 11**). The OHC-based meta-skin was affixed to the legs and feet of a movable cartoon yellow duck, creating a bionic yellow duck model (**Supplementary Movie S9**). The results revealed that the bio-inspired bioelectric response of the OHC-based meta-skin corresponded to the movement state of the bionic yellow duck, which was determined according to the rate of current change during lateral movement (**Supplementary Fig. S35**). Furthermore, we enveloped the OHC-based meta-skin around a flexible hose to fabricate a bionic garden eel model (**Supplementary Movie S10**). The results indicate that the twisting state of the OHC-based meta-skin for the bionic garden eel can be identified according to its bio-inspired bioelectric response by tracking the rate of current change during the twisting process (**Supplementary Fig. S36**). These results clearly reveal a correlation between the movement state and bio-inspired bioelectric activity. This responsiveness to physical manipulation indicates that the OHC-based meta-skin can effectively track and interpret various motions, thereby providing real-time feedback similar to that of biological tissues.

Moreover, to validate the versatility and practicality of the OHC-based meta-skin, we fabricated strips of the OHC-based meta-skin and affixed them to the surfaces of five synthetic fingers to create a model of a bionic hand. The results illustrate the accurate correspondence of bio-inspired bioelectric responses to different hand motions, such as playing basketball, carrying heavy objects, applauding, waving hands, handwriting, patting a plush toy, grabbing a mango and wiping the table; these motions show the ability of OHC-based meta-skin to replicate complex human motor functions (**Supplementary Figs. S37-44, Supplementary Movies S11-18**). By monitoring the rate of current change, the meta-skin can provide precise and reliable sensory feedback, which is crucial for the development of advanced physically intelligent materials. The above results collectively highlight the potential utility of the OHC as a valuable physically intelligent material. The adaptability of the OHC-based meta-skin to fit various bionic skeletons ensures that meta-creatures can be customized to meet specific perceptual requirements, enhancing their ability to communicate with and respond to their environment effectively and facilitating innovative applications beyond traditional robotics and prosthetics.

**OHC-based meta-systems**

The cardiovascular system is an extremely important system within an organism that is responsible for the transportation of oxygen, nutrients, and waste products[68]. In this system, arteries are responsible for conveying oxygen- and nutrient-rich blood from the heart to various parts of the body (**Fig. 5a**). Capillaries subsequently facilitate the transfer of oxygen and nutrients from the blood to tissue cells while simultaneously collecting waste products and carbon dioxide from tissue cells back into the blood. Finally, veins transport blood containing waste products and carbon dioxide from all parts of the body back to the heart. Through these mechanisms, the cardiovascular system ensures the supply of oxygen and nutrients within the organism while promoting the metabolism and elimination of waste products. Similarly, to ensure the continuous maintenance of bio-inspired biological functions in meta-creatures and explore the potential of bio-inspired systems for medical and industrial applications, a comparable transport system is imperative. Inspired by the structure and mechanism of the cardiovascular system, we developed an OHC-based meta-system that replicates key functions of the cardiovascular system, tailored for meta-creatures (**Fig. 5a**).

To compare with subsequent in vivo validation of its efficacy, we aimed to design an OHC-based meta-system that is comparable to the cardiovascular system of live rabbits. The diameter of small arteries in rabbits is approximately 0.5 mm, with a blood flow velocity of approximately 5 cm/s. The blood flow rate in these small arteries is approximately 0.6 mL/min[52]. In medium-sized arteries, the diameter is approximately 1.5 mm, with a blood flow velocity of approximately 15 cm s$^{-1}$, and the blood flow rate is approximately 16 mL min$^{-1}$. Besides, the peristaltic pump has almost no risk of damaging blood cells in maintaining blood circulation in the body. Therefore, the OHC-based meta-system, which consists of (rGO)A and (rGO)C hydrogel tubes with an inner radius of 0.7 mm and a flow rate of 12 mL min$^{-1}$, was driven by the peristaltic pump with a voltage of 12 V and a power of 5 W (**Fig. 5b**). These characteristics are intermediate between those of small arteries and medium-sized arteries.

In this system, the high-salinity solution inside two hydrogel tubes undergoes ion exchange with the low-salinity solution outside two hydrogel tubes, which is driven by osmotic pressure. The tube walls act as ion exchange membranes. The high-salinity solution inside the two tubes and the low-salinity solution outside the two hydrogel tubes are continuously drawn out, and a new solution is added to maintain a constant osmotic pressure and consistent bio-inspired bioelectricity. Thus, the OHC-based meta-system is designed to transport essential substances and remove metabolic waste in meta-creatures, thereby ensuring the continuous maintenance of their bio-inspired biological functions.

Furthermore, the results indicate that the OHC-based meta-system has a maximum current of approximately 0.15 μA and a voltage of 83 mV (**Fig. 5c**, **d**), with both parameters displaying regular fluctuations corresponding to the operation of the peristaltic pump (**Supplementary Movie S19**). Similar to the regular pumping actions performed by the heart, the peristaltic pump consistently delivers the solution, leading to periodic fluctuations in the current and voltage of the hydrogel-based cardiovascular meta-system.

Additionally, the simulations investigating electrolyte diffusion within and outside the hydrogel tube (**Fig. 5e**) corroborated the experimental findings, demonstrating effective diffusion towards the tube's outer wall and subsequent blending with the external low-concentration electrolyte solution. Notably, water from the external solution permeated through the tube wall into the interior (**Fig. 5e**), mimicking essential aspects of biological fluid transport mechanisms. These findings indicate that the OHC-based meta-system can effectively mimic the transport and metabolic functions of a biological cardiovascular system by maintaining a constant osmotic pressure and consistent bio-inspired bioelectricity; thus, the OHC-based meta-system may have application potential for use in material transport and waste metabolism.

**Application of the OHC-based meta-system in living organisms**

To evaluate the integration potential of the OHC-based meta-system into living organisms and prevent rabbit mortality, we adopted an open-source approach, preventing blood from re-entering the rabbit's body. To observe the blood flowing through the hydrogel tube, we did not add rGO materials when preparing the transparent hydrogel tube. The originally separated A and C hydrogel tubes served different functions and were combined into a single hydrogel tube as an integrated OHC-based hydrogel tube (**Fig. 5f**). The integrated OHC-based meta-system, which consists of an integrated OHC-based hydrogel tube with an inner radius of 0.7 mm and a flow rate of 12 mL min$^{-1}$, was driven by a voltage of 12 V and a power of 5 W (**Supplementary Movie S20**). These features also fall between small- and medium-sized arteries, ensuring their compatibility with the blood flow velocity in rabbits.

Here, the rabbits weighed 3 kg (whole blood volume: 55–70 mL kg$^{-1}$). Moreover, the myocardial potential changes in rabbits range from approximately -90 to +20 mV, and the action current of their myocardial cells ranges from a few nanoamps (nA) to several tens of nanoamps[53]. Concurrently, each rabbit was injected with an appropriate amount of saline solution to dilute its blood concentration, thereby avoiding death due to excessive blood loss. Furthermore, we use an infusion needle to puncture the rabbit auricular artery and introduce its blood through a matching hose to the integrated OHC-based meta-system (**Fig. 5f**). Finally, the blood exits this system and enters a Petri dish that collects waste blood.

Moreover, the results indicate that as blood flows through the integrated OHC-based meta-system, it generates currents and voltages ranging from approximately 0.10 to 0.16 μA and 40 mV, respectively (**Fig. 5g**). Notably, these systems are close to those associated with rabbits; thus, the integrated OHC-based meta-system could respond dynamically to biological fluids, maintaining functionality under physiological conditions. The observed fluctuations in current and voltage further highlight the system's ability to adapt to varying flow rates and pressures within living tissues.

**Rheological mechanism of the OHC-based meta-system**

To elucidate rheological properties of the integrated OHC-based meta-system, we simulated the pattern of pulsatile flow generated as blood passes through a peristaltic pump (**Supplementary Text 12**). The results indicate that the roller begins to enter the compression phase at t = 0.3 s, progressively increasing the pressure exerted on the tube (**Supplementary Fig. S45**). As the fluid space within the tube gradually decreases, the fluid is expelled through both the inlet and outlet. At t = 0.5 s, the compression effect of the roller reaches its maximum (**Supplementary Fig. S46**). As the roller moves upward along the tube, the fluid follows suit, moving upwards at both the inlet and the outlet. This movement is where most of the net flow from the inlet to the outlet is generated. Finally, at t = 1.3 s, the compression process begins to reverse as the roller disengages from the tube; as a result, fluid flows into the tube from both ends. The flow conditions at the inlet and outlet indicate that the peristaltic pump generates volume changes, which in turn cause the fluid to move forward in a pulsatile manner (**Supplementary Fig. S47**). This pulsatile flow induces pulse-like deformation and vibration in the hydrogel tube, closely resembling the pulsatile nature of blood flow generated by the heart.

Furthermore, we simulated the viscoelastic flow of blood through the soft hydrogel tube of the integrated OHC-based meta-system. The study of viscoelastic flow utilizes the Weissenberg number to compare the magnitudes of elastic and viscous forces (**Supplementary Text 13**). In our model, we employed three different hydrogel polymer relaxation times: 0.05, 0.1, and 0.2 s, corresponding to Weissenberg numbers $Wi = 0.15$, $Wi = 0.3$, and $Wi = 0.6$, respectively. The results indicate that at a Weissenberg number of 0.6, the deformation of the hydrogel tube and the boundary loads on the hydrogel tube wall are illustrated in **Supplementary Fig. S48**. Upward arrows represent loads generated due to interaction with the fluid, while downward arrows indicate external pressure on the elastic hydrogel tube. The normal force on the tube wall is primarily caused by the pressure on the elastic hydrogel tube (**Supplementary Fig. S49**). The pressure from the fluid on the hydrogel tube acts upward, opposite to the direction of external pressure, but is smaller, resulting in downward deformation of the hydrogel tube (**Supplementary Fig. S50**). Although the normal component of viscoelastic stress does not directly affect the hydrogel tube, different relaxation times lead to varying distributions of pressure and hydrogel tube displacement. Increasing the Weissenberg number results in higher pressure downstream in the narrowest gap of the channel and slightly reduces the hydrogel tube displacement. The narrowing of the channel also affects the velocity distribution (**Supplementary Fig. S51**). These results suggest that the flowing blood exerts pressure on the inner surface of the hydrogel tube, which deforms the hydrogel tube, and then changes the magnitude of the bio-inspired bioelectricity of the integrated OHC-based meta-system.

In summary, these findings collectively validate the use of OHC-based meta-systems as promising tools for mimicking and potentially enhancing natural cardiovascular functions. By maintaining osmotic pressure and consistent bio-inspired bioelectric performance similar to that of biological systems, these meta-systems have significant promise for applications in material transport, waste metabolism, and potentially bio-integrated technologies in meta-creatures and living organisms. Further research and development in this area could facilitate innovative biomedical and industrial applications leveraging bio-inspired design principles.

**Conceptual model of the meta-creature**

Finally, inspired by the symbiotic relationship between the embryos of spotted salamanders and algae capable of photosynthesis, we developed a bionic bird model that integrated an OHC-based meta-system, conceptualized as a meta-creature, to assess the application potential of OHC-based meta-systems in meta-creatures (**Fig. 5h**). Here, the OHC-based meta-system, which consists of (rGO)A and (rGO)C hydrogel tubes with an inner radius of 0.7 mm and a flow rate of 12 mL min$^{-1}$, was driven by a voltage of 12 V and a power of 5 W (**Fig. 5h**). The OHC-based meta-system is engineered to perform crucial functions throughout the meta-creature's body, maintaining physiological activities similar to those in biological systems. These functions include the transport of essential substances and the removal of metabolic waste, facilitated by the ion exchange driven by osmotic pressure within the system. The tube walls of the OHC-based meta-system act as ion exchange membranes, ensuring consistent bio-inspired bioelectricity and maintaining osmotic pressure; this aspect is vital for sustaining bio-inspired bioelectric functionality.

To evaluate its potential for application, we conducted flight tests with the meta-creature held at a fixed position, simulating outdoor flight conditions (**Fig. 5h**). The experimental results indicate that this design allows the meta-creature to generate bio-inspired bioelectricity during simulated natural flight conditions (**Supplementary Movie S21**). During flight, the amplitude current generated by the wing vibration of the metamaterial fluctuates within 0–10 μA (maximum of approximately 24 μA), and the amplitude voltage fluctuates within 0–4 mV (maximum of approximately 9 mV) (**Fig. 5i**, **j**). During the wing vibration, the peak amplitude current (approximately 10 μA) is significantly greater than the current (maximum of approximately 0.15 μA) of the aforementioned OHC-based meta-system and the current (approximately 0.10 to 0.16 μA) of the OHC-based meta-system applied in rabbits; however, the peak amplitude voltage (approximately 4 mV) is lower than those voltages. These results occur because we performed first derivative operations on the original current–voltage–time relationships to better display the principles of bio-inspired bioelectricity during flight. These bio-inspired bioelectricity signals indicate that the OHC-based meta-system integrated into the meta-creature effectively mimics the cardiovascular functions of natural creatures. Furthermore, these findings suggest that the compatibility of the OHC-based meta-system with the meta-creature extends beyond mere functionality, each component of the meta-creature is then provided with sensory capabilities and the ability to achieve biological activity levels comparable to those observed in natural organisms.

## Operating principle of the OHC-based meta-system during the meta-creature's flight

To reveal the operating principle of the fluid passing through the OHC-based meta-system during the process of the meta-creature vibrating its wings, we simulated the interaction between the fluid passing through the elastic bending pipe and the movement of the pipe during vibration (**Supplementary Text 14**). The results indicate that the cross-sectional view of the velocity filed at an inlet mass flow rate of 0.2 kg s$^{-1}$ (**Supplementary Fig. S52**), suggests that the fluid exerts pressure on the pipe walls as it flows through the bend. The pressure field exerted by the fluid on the inner wall of the bend is depicted in **Supplementary Fig. S53**. These findings demonstrate that as the fluid enters the pipe in a specified direction, the curvature of the pipe alters the fluid's established flow path, exerting significant pressure on the curved surface that redirects the flow. During the vibration of the bend, this interaction may also cause substantial displacement of the unrestrained pipe (**Supplementary Fig. S54** and **Supplementary Movie S22**). Consequently, in this study, this interaction leads to deformation during the vibration of the restrained bend, causing variations in current within the meta-creature that are consistent with the wing vibration pattern.

## Discussion

In pursuit of creating life-like creatures, we introduced a groundbreaking strategy utilizing omnipotent hydrogel cells (OHCs) as meta-cells, enabling the manufacture of meta-nerve fibres, meta-skins, meta-mouths, and meta-cardiovascular systems through the ingenious functions of disintegration, reconfiguration, and regeneration at the meta-cell level. The conceptual model of a meta-creature emerged from this intricate process. The core of our approach is the utilization of free radical polymerization, ion-selective diffusion, and bio-inspired bioelectricity signals. These mechanisms enable the replication of essential life processes at the meta-cell level, and these processes include metabolism, proliferation, regeneration, raw material intake, self-regulation, reconfiguration, repairability, deformation, and stress response, along with their corresponding bio-inspired bioelectricity signals. A key innovation of our work is in the design and validation of a meta-cardiovascular system capable of material transportation and waste metabolism, exemplified by successful testing on live rabbits. Remarkably, our meta-creature, which is equipped with a meta-cardiovascular system, generated bio-inspired bioelectricity signals during simulated flight, highlighting its potential for practical applications. This research contributes a novel biocultural evolution theory centred on the potential birth of technological life, advancing our understanding of the bottom-up evolution of OHCs into living-like entities and providing a framework to explore deep future models of such life forms. While the attainment of life-like processes and their corresponding bioelectricity by meta-cells currently relies on human control in experiments, our proposed strategy has theoretical promise in establishing bio-like functions.

To enable the application of OHC and its derivatives in living creatures, we designed the OHC primarily composed of carbon, hydrogen, oxygen, and nitrogen because these elements are also predominant in biological proteins. Since the generation of electrical signals in biological cells[54] and the generation of bio-inspired electrical signals in OHC largely depend on ion concentration gradients, maintaining these signals over a prolonged period is challenging. Despite our efforts to achieve relatively stable bio-inspired bioelectric output from OHC within a limited timeframe using methods such as the incorporation of reduced graphene oxide, metabolism, intake of raw materials, proliferation, and regeneration, we were unable to consistently overcome the fundamental limitation imposed by the diffusion of ion concentration gradients[55].

However, while investigating the regeneration of OHCs, we unexpectedly discovered that each component of the OHC (H, (rGO)C, L, (rGO)A) could be regarded as generalized OHC hydrogel cells capable of generating bio-inspired bioelectric signals. Furthermore, we found that OHCs could decompose, reconstruct, and regenerate at multiple scales; this enabled them to perform the following functions: adapt to changes in shape and size; adjust their mechanical properties; control their current and voltage; assemble and program their bionic functions; and transform their biomimetic functionalities. Notably, this disintegration-reconstruction-regeneration mechanism is effective even at the microscopic level, facilitating the construction of micro-biosystems and artificial devices.

Notably, we found that the reconstructed OHCs contain numerous series-parallel structures and that the number of monomers after disintegration determines the initial discharge intensity of the reconstructed units. This insight indicate that we can temporarily increase the energy output of meta-systems within a short time frame through the disintegration-reconstruction method, greatly expanding their application potential. Initially, the combat capabilities of meta-systems are significantly enhanced. The speed and agility of meta-systems is also improved. Additionally, due to their potential activation, meta-systems may unleash special abilities that are otherwise difficult to activate, such as energy projection[56] or environmental manipulation[57]. In extreme cases, the high-energy projection may even influence the intelligent systems of other bionic creatures[58].

Moreover, by utilizing parallel and series configurations, size adjustments, or integration, the magnitude of bioelectricity in OHCs can be effectively controlled, thereby broadening the potential applications of OHCs and OHC-based devices in biomedical engineering, biomedical devices, and engineering technologies. However, our results indicate that although extensive series connections can achieve very high voltages, parallel connections are not an effective method for obtaining high current. Some OHCs in parallel configurations exhibit capacitor-like characteristics[40-42], which hinder the increase in current in parallel setups. Compared with the series method, the disintegration-reconstruction approach can quickly and efficiently increase the output with relatively fewer OHCs, thereby reducing material waste. In the future, we may focus more on achieving high current outputs for bionic equipment based on OHCs rather than merely increasing the voltage through large-scale series connections.

Furthermore, the successful development of 2.4-metre-long OHC-based meta-nerve fibres provides the foundation for creating neural systems in macroscopic meta-creatures, enabling the establishment of complex and extensive bio-inspired neural networks. However, the relationship between the bio-inspired bioelectric signals generated by these bionic nerves and the behaviour of meta-creatures may be complex and requires collaborative exploration from researchers across various disciplines. Additionally, while the designed meta-mouth and meta-skin enable these meta-creatures to experience and respond to regular environmental stimuli in a manner similar to humans (such as specific liquid concentrations, playing basketball, or repeatedly lifting heavy objects), the correlation between these bio-inspired bioelectric signals and irregular environmental stimuli needs to be thoroughly investigated.

To sustain the biomimetic functions of meta-creatures over time, we drew inspiration from the structure and mechanisms of the cardiovascular system. We tailor a meta-system based on OHCs, allowing it to effectively mimic the transport and metabolic functions of biological cardiovascular systems by maintaining constant ion concentration gradients and bioelectric stability. However, the hydrogel tubes constituting the meta-system swell over prolonged immersion in aqueous solutions, affecting the stability of its performance. In the future, more research is needed to overcome this challenge. Although we demonstrated the integration potential of the OHC-based meta-system with the cardiovascular system of living rabbits, the system remains an open-source approach since we currently lack the capability to circulate rabbit blood through the meta-system and return it to their bodies. Moving forwards, further exploration is needed to investigate methods for integrating meta-systems with the cardiovascular systems of living organisms.

Finally, we demonstrated that the OHC-based meta-system integrated on a bionic bird effectively provides each component of the meta-creature with sensory capabilities. However, interpreting complex, multisite bioelectric response relationships remain a daunting challenge. Thus, collaborative efforts may be needed employing intelligent algorithms[59], big data model construction[60], machine learning[61], and other methods. While advancements in artificial intelligence[62] may soon enable the attainment of meta-creatures, their potential applications also cause numerous uncertainties and risks. Therefore, a cautious approach is crucial in the development of meta-creatures to steer them towards directions beneficial for the advancement of human civilization.

# Acknowledgements

We acknowledge this publication is supported by and coordinated through Zhongshan Hospital of Fudan University and the Shanghai Key Laboratory of Craniomaxillofacial Development and Diseases (Qiangqiang Zhou and Xiaoling Wei).

**Funding:** This work was supported by the National Natural Science Foundation of China (62305068 and 62074044), China Postdoctoral Science Foundation (2022M720747), Shanghai Post-doctoral Excellence Program (2021016), Shanghai Rising-Star program (22YF1402000), and Zhongshan-Fudan Joint Innovation Center and Jihua Laboratory Projects of Guangdong Province (X190111UZ190).

**Author contributions:** H.Q.D., W.Q.D., G.Q.Z. and R.Q.G. conceived and designed the project. H.Q.D. and Y.Y.C. generated the data. H.Q.D. and W.Q.D. analysed and interpreted the data. H.Q.D. and W.Q.D. wrote the manuscript. G.Q.Z. and R.Q.G. supervised the study. Other authors assisted in the completion of this project. All authors edited and approved the manuscript.

**Competing interests:** Authors declare that they have no competing interests.

**Supplementary Materials**

Supplementary Texts 1-14

Methods 1-25

Figs. S1 to S54

Tables S1 and S2

Supplementary Movies S1-S22

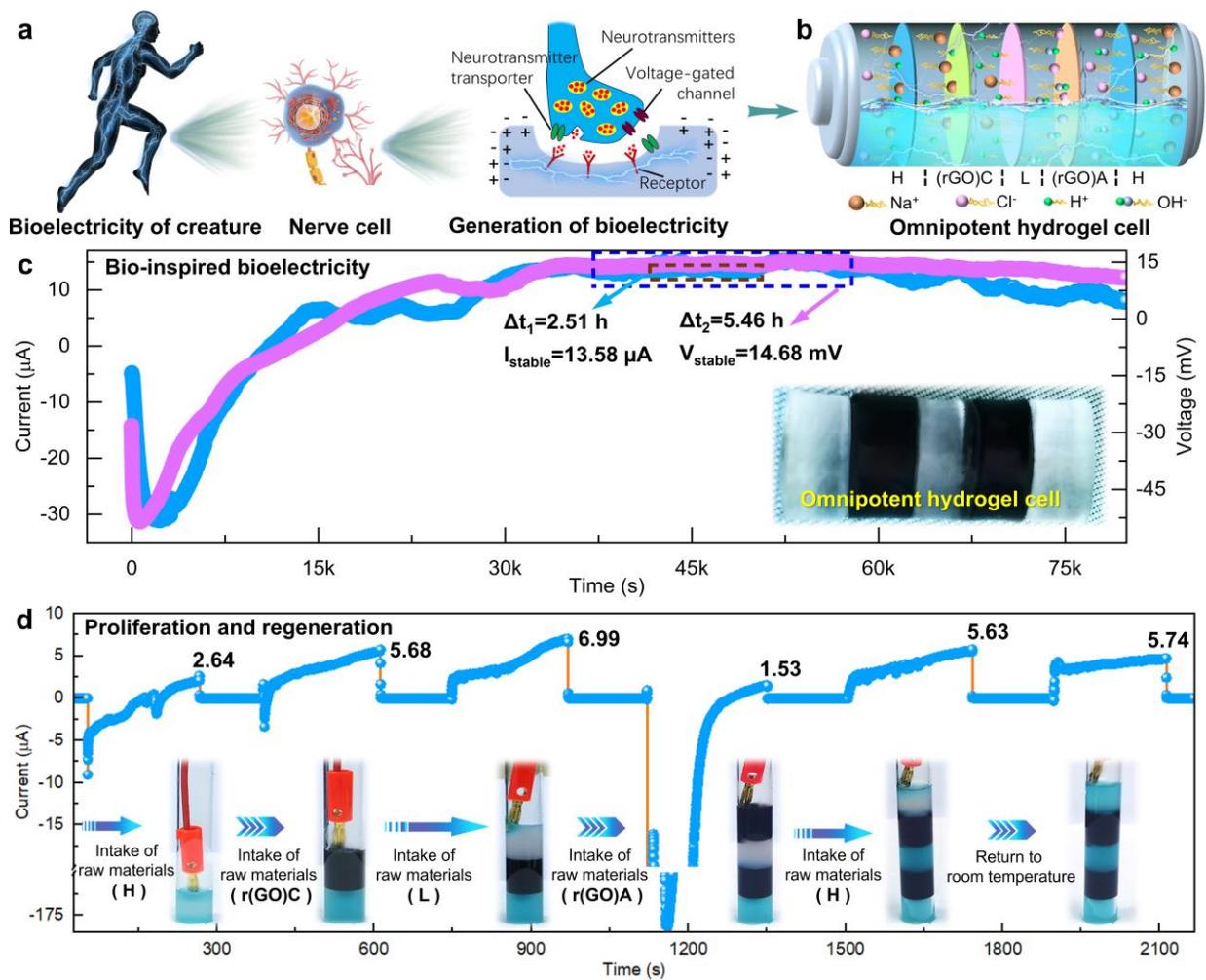

**Fig. 1. Design strategy, bio-inspired bioelectricity mechanism and properties of the OHC.**
**a**, Diagram of the OHC inspired by bioelectricity. **b**, Diagram of the bio-inspired bioelectricity mechanism and structure of the OHC. **c**, Bio-inspired bioelectricity of the OHC at room temperature. The illustration is a photo of the OHC. **d**, Proliferation and regeneration processes of the OHC at a visible colour change rate.

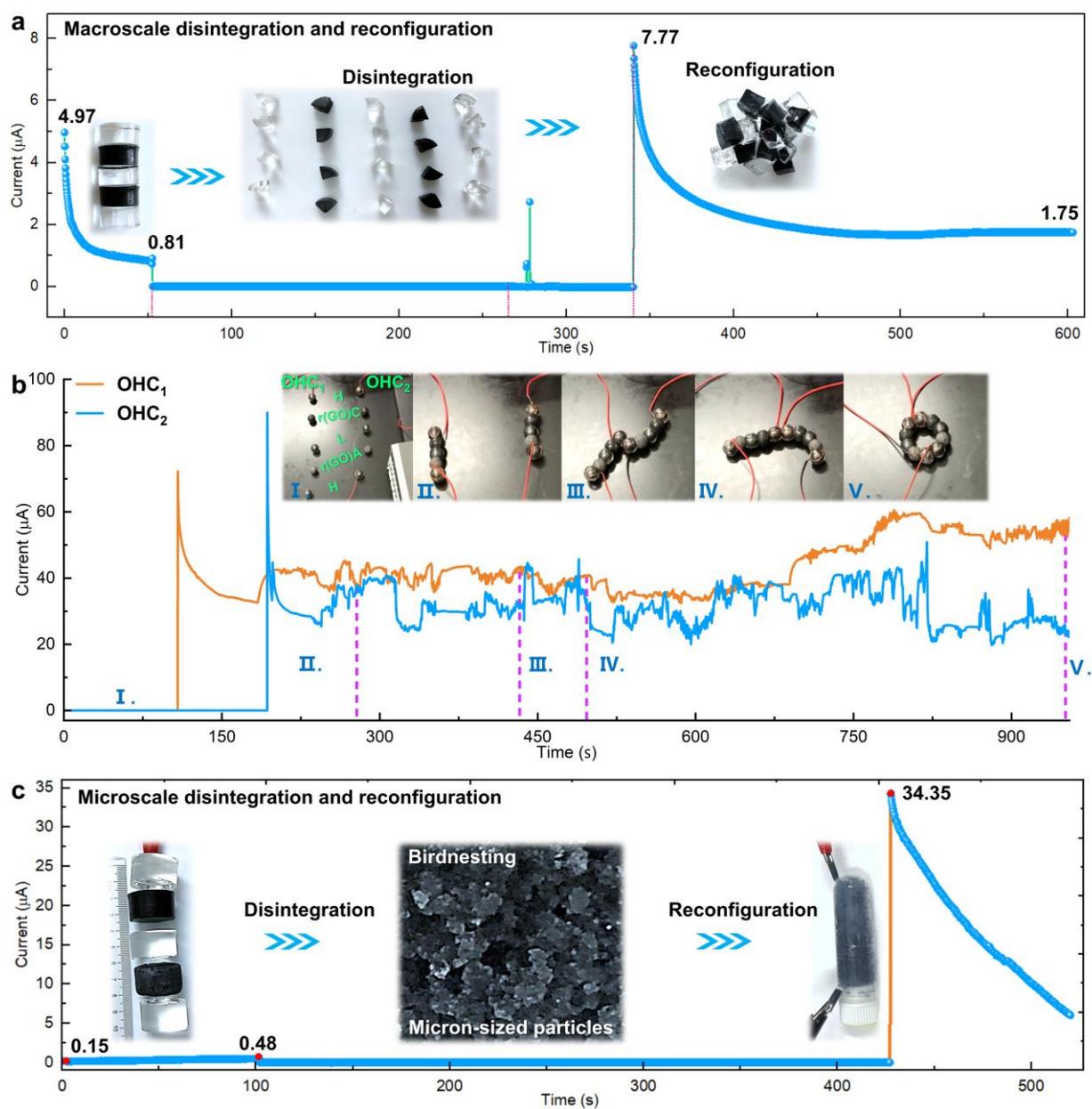

**Fig. 2. Disintegration-reconfiguration properties and activity of the omnipotent hydrogel cell.** The bio-inspired bioelectricity of the disintegration-reconfiguration capability of the OHC at the macroscale (**a**) and the microscale (**c**). **b**, Activities of disintegration and reconfiguration for two OHCs. The illustrations show their specific activities.

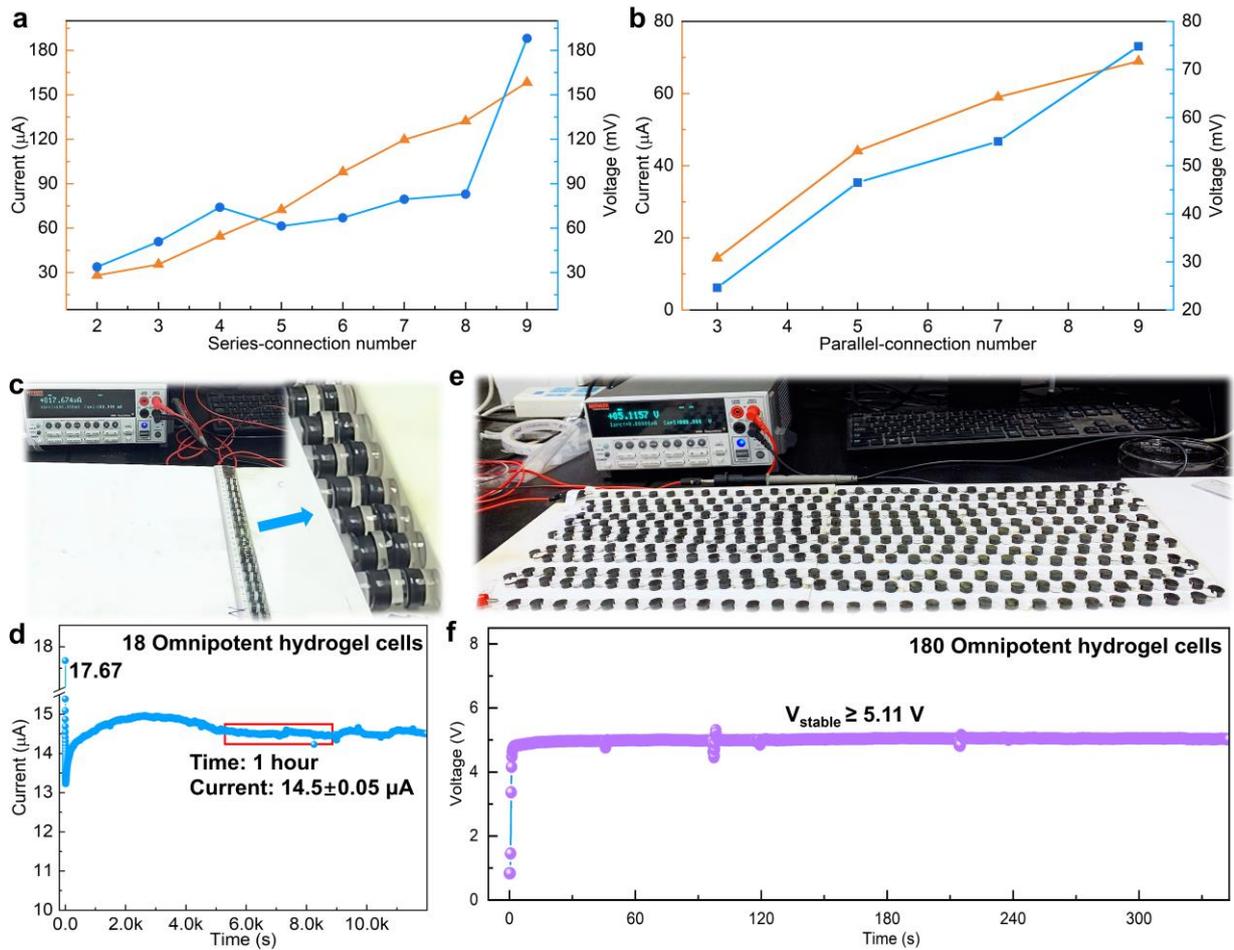

**Fig. 3. Bio-inspired bioelectricity of the ordered series and parallel connections for OHCs.**
**a,** Bio-inspired bioelectricity of the ordered series connections with the number of connections ranging from 2 to 9. **b,** bio-inspired bioelectricity of the ordered parallel connections with 3, 5, 7 and 9 connections. **c,** Photo of 18 OHCs in parallel. The illustration shows the specific connection mode, and the diameter and thickness of each gel were 1 cm and 0.5 cm, respectively. **d,** Bio-inspired bioelectricity of the ordered series connections for 18 OHCs. **e,** Photograph of 18 OHCs in parallel and their specific connection mode. **f,** Bio-inspired bioelectricity of the ordered parallel connections for 180 OHCs.

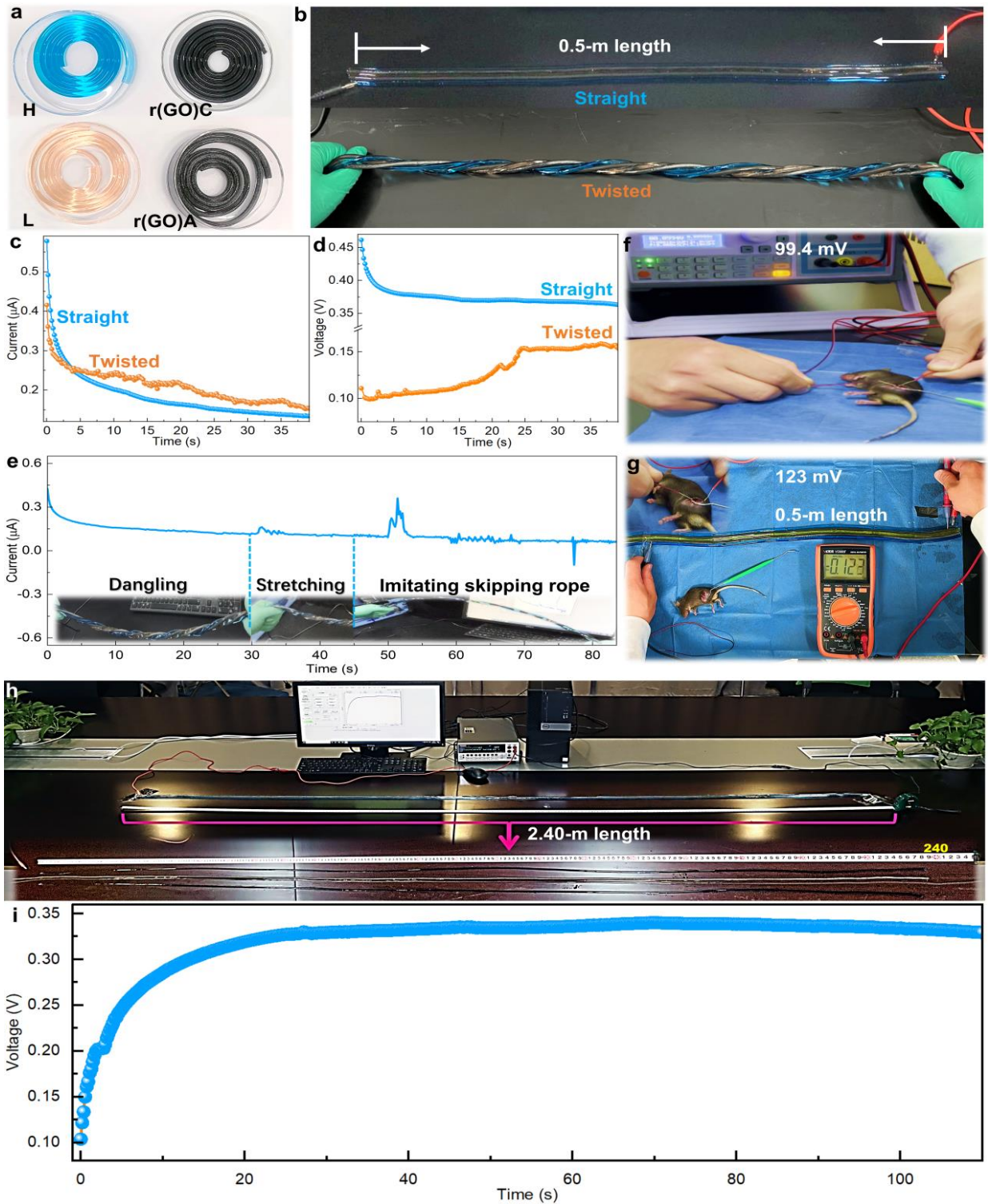

**Fig. 4. Bio-inspired bioelectricity of OHC-based bio-inspired nerve fibres. a**, Photos of the 0.5-m long H, r(GO)C, L and r(GO)A hydrogel fibres. **b**, Photos of the 0.5-m long OHC-based bio-inspired nerve fibres in the straight and twisted states. **(c, d)**, Bio-inspired bioelectricity of the 0.5-m length OHC-based bio-inspired nerve fibre in the straight and twisted states. **e**, Bio-

inspired bioelectricity of the 0.5-m long OHC-based bio-inspired nerve fibre in the dangling, stretching and imitating rope skipping states. **f,** Sciatic nerve of a rat stimulated by a voltage output source metre with a voltage of 99.4 mV. **g,** Sciatic nerve of a rat stimulated with the 0.5-m long OHC-based bio-inspired nerve fibre. **h,** Photo of the 2.4-m long OHC-based bio-inspired nerve fibre during a test of bio-inspired bioelectricity. **i,** Bio-inspired bioelectricity of the 2.4-m long OHC-based bio-inspired nerve fibre.

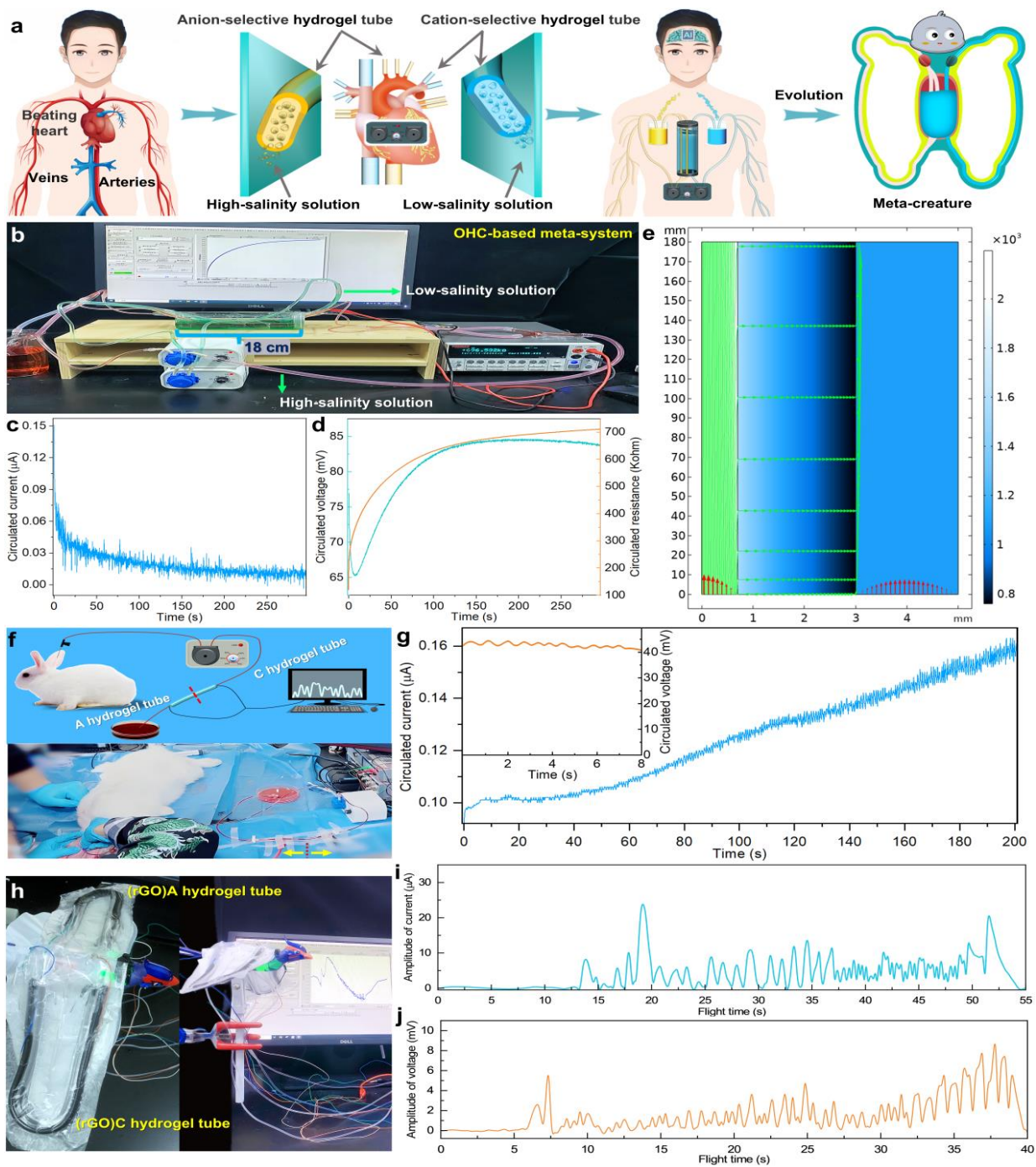

**Fig. 5. Design, performance and applications of the OHC-based meta-system. a,** Photo of a conceptual model for meta-creation with the hydrogel-based cardiovascular meta-system. **b**, Photo of the OHC-based meta-system. (**c**, **d**), Bio-inspired bioelectricity of the OHC-based meta-system. **e,** Finite element analysis of the solute diffusion for the OHC-based meta-system. **f,** Schematic diagram and photo of the OHC-based meta-system integrated into one hydrogel tube

connected to the rabbit auricular artery. The inner radius of both tubes is 0.7 mm, and the thickness of both tubes is 2.3 mm. The A hydrogel tube is prepared from a cation-selective hydrogel. The C hydrogel tube is prepared from an anion-selective hydrogel. **g,** Bio-inspired bioelectricity of the OHC-based meta-system integrated into one hydrogel tube connected to the rabbit auricular artery. **h,** Photo of the meta-creature conceptual model and integrated OHC-based meta-system. The inner radius of both tubes of the integrated OHC-based meta-system is 0.7 mm, and the thickness of both tubes is 2.3 mm. The (rGO)A hydrogel tube is prepared from the rGO composite cation-selective gel. The (rGO)C hydrogel tube is prepared from the rGO composite anion-selective gel. (**i, j**), Bio-inspired bioelectricity of the meta-creature conceptual model.